# Giant Nonreciprocity of Surface Acoustic Waves induced by a positive-negative magnetostrictive heterostructure

Wenbin Hu[1,2], Mingxian Huang[1,2], Yutong Wu[3], Yana Jia[3], Wen Wang[3], and Feiming Bai[1,2]*

[1]School of Electronic Science and Engineering, University of Electronic Science and Technology of China, Chengdu, 610054, China

[2]State Key Laboratory of Electronic Thin Films and Integrated Devices, University of Electronic Science and Technology of China, Chengdu 610054, China

[3]Institute of Acoustics, Chinese Academy of Sciences, Beijing, 100190, China

* To whom correspondence should be addressed. Electronic mails:

fmbai@uestc.edu.cn

**Abstract**

Lack of nonreciprocity is one of the major drawbacks of solid-state acoustic devices, which has hindered the development of microwave-frequency acoustic isolators and circulators. Here we report giant nonreciprocal transmission of shear-horizontal surface acoustic waves (SH-SAWs) on a LiTaO$_3$ substrate coated with a negative-positive magnetostrictive bilayer structure of Ni/Ti/FeCoSiB. Although the static magnetic moments of two layers are parallel, SH-SAWs can excite optical-mode spin waves much stronger than acoustic-mode ones at relatively low frequencies via magnetoelastic coupling. The measured magnitude nonreciprocity exceeds 40 dB (or 80 dB/mm) at 2.333 GHz. In addition, maximum nonreciprocal phase accumulation reaches 188° (376°/mm), which is desired for an effective SAW circulator. Our theoretical model and calculations provide an insight into the observed phenomena and demonstrate a pathway for further improvement of nonreciprocal acoustic devices.

Owing to the significantly shorter wavelengths of acoustic waves compared to electromagnetic waves (EMWs) and the energy constraint in solid media, surface acoustic wave (SAW) filters have been miniaturized and achieved very high quality factors, leading to tremendous success in the field of RF signal processing. The high efficiency of SAW is particularly attractive for device applications[1-3]. Recent research has demonstrated that SAW can excite magnon-phonon coupling in ferromagnetic thin films, enabling long-distance propagation of spin waves[4] (SWs) and non-reciprocal transmission of SAWs. The latter arises from the helicity mismatch between Rayleigh-type SAWs and the chirality of the magnetization precession in ferromagnetic thin films[5-9], as well as the interfacial Dzyaloshinskii-Moriya interaction (iDMI) between ferromagnetic thin films and heavy metals[10,11]. However, the nonreciprocity caused by helicity mismatch is relatively weak. iDMI is generally accompanied by large magnetic damping due to spin pumping effect, resulting in severe insertion loss.

Nonreciprocal transmission of SAW has also been reported in bilayer structures composed of ferromagnetic metal ($FM_1$)/non-magnetic spacer (NM)/ferromagnetic metal ($FM_2$). In these structures, the dispersion relation of SWs is nonreciprocal due to the break of the space-time inversion symmetry via interlayer dipolar interaction[12,13]. When both the resonance frequency and wavevector of SAWs match those of SWs, intense phonon-magnon coupling occurs with significant power absorption due to magnetization precession, meanwhile, the propagation loss of SAWs remains low when SAWs and SWs are uncoupled. Recently, researchers have utilized this feature to achieve large nonreciprocal SAW transmission in NiFe/Au/CoFeB[14], $FeGaB/Al_2O_3/FeGaB$[15,16], NiFeCu/FeCoSiB[17] bilayer and synthetic antiferromagnetic (SAFM) structures of Co/Ru/Co[18,19] and CoFeB/Ru/CoFeB[20]. However, SAFM requires precise control of the thickness of the space layer over a large area. In addition, the requirements of IDC and SAFM on the thickness of magnetic layers are different. While high IDC prefers a large magnetic layer thickness and wavenumber, the antiferromagnetic coupling becomes weaker with increasing layer thickness. Therefore, most SAWs devices based on SAFM typically achieve strong nonreciprocity at large wavenumbers or high frequencies of 5-8 GHz, much higher than commercial SAWs devices (below 3 GHz).

In this work, we employed a negative-positive magnetostrictive bilayer structure of Ni/Ti/FeCoSiB with negative and positive magnetostriction to investigate the transmission characteristics of SAWs. Distinct from EMWs, SAWs can excite strong optical-mode SWs in Ni/Ti/FeCoSiB via magnetoelastic coupling (MEC) at relatively low frequencies, although the static magnetization of two layers are parallel with each other. The negtiave-positive magnetostriction configuration can greatly enhance the nonreciprocal transmission of SAWs.

Fig. 1(a) illustrates the structure of the SAW delay line and our experimental setup. A 42° -rotated Y-cut X-propagation $LiTaO_3$ substrate was selected to excite shear horizontal type SAWs (SH-SAWs). Interdigital transducers (IDTs) of Ti (5 nm)/Al (50 nm) were deposited on a $LiTaO_3$ substrate via sputtering, and the spacing between two IDT pairs is 600 μm. A split-finger design with a pitch width and interval of 2.4 and 1.6 μm was employed to suppress the reflection of SAWs and obtain high-order harmonics[21]. Each IDT has 5 pairs of fingers. A 0.6×0.5 mm² rectangular structure of Ni(16 nm)/Ti(8 nm)/FeCoSiB(16 nm)/Ti(10 nm) is then deposited and patterned between the two IDTs. The bottom Ti layer serves as a seed layer to facilitate the subsequent layer growth. The central Ti layer separates the Ni and FeCoSiB layers, thereby inhibiting interlayer exchange coupling. Fig. 1(b) shows the photograph of our fabricated device and the coordinate system with the *x*-axis parallel to the SAW propagation direction and the *z*-axis perpendicular to the plane of the magnetic films. Here, $\varphi_M$ and $\varphi_H$ are the angles between the magnetic moment and the applied field with respect to the *x*-axis. During the sputtering process of both FeCoSiB and Ni layers, an in-situ magnetic field about 150 Oe was applied along the *y*-axis to induce uniaxial anisotropy, hence $\varphi_M$ is 90° under zero magnetic field.

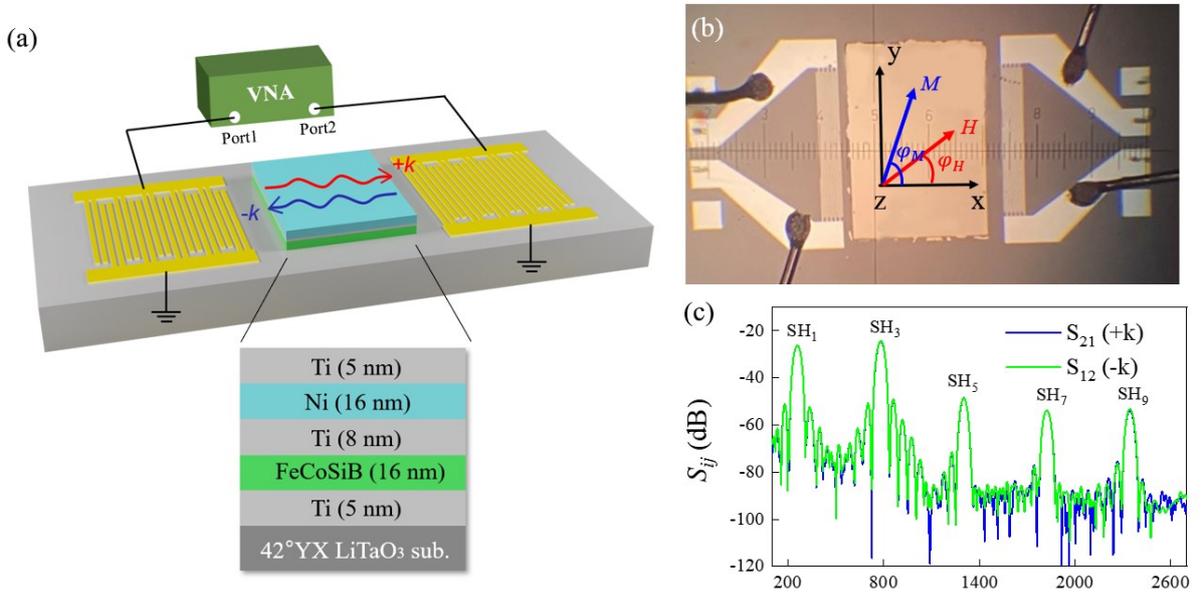

FIG. 1 (a) Schematic illustration of a SH-SAW delay line based on a Ni(16 nm)/Ti(8 nm)/FeCoSiB(16 nm) heterostructure on a LiTaO$_3$ substrate. (b) Optical image of the SAW delay line. (c) Measured SAW transmission parameters upon applying a fixed magnetic field of 300 Oe after time-domain gating.

Two-port transmission parameters of the delay lines were then measured by a vector network analyzer (VNA, Agilent N5230A). After performing time-domain gating, we can obtain the forward ($S_{21}$) and backward ($S_{12}$) transmission characteristics of the SAWs passing through the films. Although the fundamental resonant frequency of SH-SAWs is 257 MHz, high odd-order harmonic modes up to the 9th order (2333 MHz) are present due to the split-finger design of IDTs. The SAW wave vector $k_{SAW}$ can be obtained using the formula $|k_{SAW}| = 2\pi f_{SAW}/v_{SAW}$, where $v_{SAW}$ is SH-SAW velocity (~4112 m/s). Fig. 1(c) shows a wideband measurement of $S_{21}$ and $S_{12}$ from 100 to 2700 MHz under an applied field of 300 Oe. It can be seen that $S_{21}$ almost overlaps with $S_{12}$, indicating the absence of nonreciprocity under this field.

We employ a phenomenological model (see Section A of Supplementary Material) to describe the SAW-SW coupling in the magnetic bilayers structure. The effective fields in the model include the external magnetic field, the in-plane uniaxial anisotropy field, the dipolar field, as well as the intralayer exchange interaction field. Due to the relatively thick space layer between the two FM layers, the interlayer exchange coupling was ignored.

Fig. 2(a) shows the calculated SW dispersion curves for Ni (16 nm), FeCoSiB (16 nm) and Ni (16 nm)/Ti (8 nm)/FeCoSiB (16 nm). For a single-layered magnetic film like Ni or FeCoSiB, symmetrical dispersion curves are observed. The spin wave resonance (SWR) frequency of Ni is much lower than that of FeCoSiB due to its lower saturation magnetization $M_s$. However, the SW dispersion curves of the Ni/Ti/FeCoSiB bilayer include both in-phase (namely acoustic mode, AM) and out-of-phase (namely optical mode, OM) branches. Of particular interest is that the OM frequency is lower than that of the AM, even lower than those of either Ni or FeCoSiB layers[14,22]. For $|k| \leq 6 \, \mu m^{-1}$, the calculated OM frequencies are below 2.5 GHz, which benefits the coupling between SAWs and SWs. Additionally, the SW dispersion curves of both modes are nonreciprocal, which can be attributed to the interlayer dipolar interaction mentioned above. Fig. 2(b) depicts the OM dispersion curves for different values of $\varphi_M$. We define frequency offset of OM at $\pm k$ as $\Delta f_{OM} = f_{OM}(+k) - f_{OM}(-k)$, which gradually decreases when $\varphi_M$ changes from 90° to 0° and vanishes at $\varphi_M = 0°$. Therefore, the largest nonreciprocity of spin wave appears at $\varphi_M = 90°$.

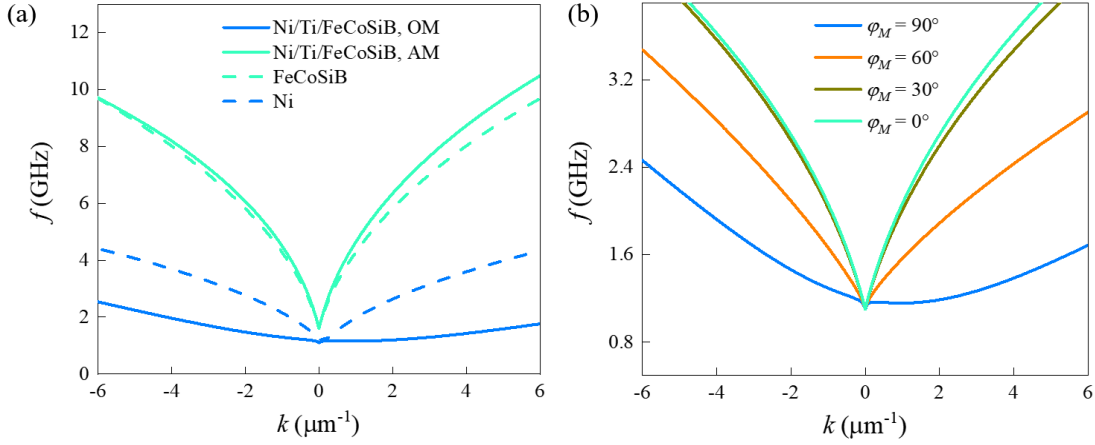

FIG. 2 (a) Calculated spin wave dispersion curves of Ni (16 nm), FeCoSiB (16 nm) and Ni (16 nm)/Ti (8 nm)/FeCoSiB (16 nm) at zero bias field. (b) Optical-mode dispersion curves for different $\varphi_M$ angles.

As reported in previous studies, SWR can be excited by SAWs via MEC. For the SH-type SAWs discussed in this work, the main strain component within the magnetic layer is $\eta_{xy}$ and the effective driving field $\boldsymbol{h}^X$ can be denoted as

$$\begin{pmatrix} h_\theta^X \\ h_\varphi^X \end{pmatrix} = -\frac{\eta_{xy}}{\mu_0 M_s^X} \begin{pmatrix} 0 \\ 2B_2^X \cos(2\varphi_M^X) \end{pmatrix}, \qquad (2)$$

where $B_2^X$ is magnetoelastic coupling constant in layer $X$ ($X = A$ or $B$), and $(\theta, \varphi)$ are spare angles in a spherical coordinate system. Clearly, the effective driving field of SH-SAW exhibits a $cos(2\varphi_M)$ angle dependency. In other words, the strongest MEC is observed at 90°. Additionally, the in-plane interlayer dipolar field in the bilayer structure is also associated with $\varphi_M$ and can be written as

$$\begin{aligned} h_\varphi^{IDC,A} &= M_s^B \frac{d^B}{2} G^A G^B e^{-|k|s} (-i \sin \varphi_M^A k \delta \theta^B + \sin \varphi_M^A \sin \varphi_M^B |k| \delta \varphi^B), \\ h_\varphi^{IDC,B} &= M_s^A \frac{d^A}{2} G^A G^B e^{-|k|s} (i \sin \varphi_M^B k \delta \theta^A + \sin \varphi_M^A \sin \varphi_M^B |k| \delta \varphi^A) \end{aligned} \qquad (3)$$

with $G^X = (1 - e^{-|k|d^X})/|k|d^X$. The $i\sin\varphi_M^X k\delta\theta^X$ term contributes to the SW nonreciprocity, which maximizes with $\varphi_M = 90°$, consistent with the computational results shown in Fig. 2(b). Therefore, SH-SAW is more proper to generate larger nonreciprocal characteristics compared to well-studied Rayleigh wave[23,24].

When the frequency and wavenumber of SAWs match those of SWs, $\boldsymbol{h}^X$ can excite a resonantly enhanced magnetic precession $M_s^X \delta \boldsymbol{m}^X$, which results in the absorption of SAWs power. The total power absorption $P_{abs}$ can be expressed as the sum of absorbed powers in each ferromagetic layer and denoted as

$$P_{abs} = \pi f \mu_0 W \int_0^L \text{Im}(M_s^A (\boldsymbol{h}^A)^* \delta \boldsymbol{m}^A + M_s^B (\boldsymbol{h}^B)^* \delta \boldsymbol{m}^B) dx, \qquad (4)$$

where $W$ refers to the IDT aperture and $L$ denotes the length of the ferromagnetic layer. Using Eq. (4), we calculated the $P_{abs}$ of three bilayers with the same bottom layer of FeCoSiB (16 nm) but different upper layers of Ni (16 nm), $Ni_{81}Fe_{19}$ (16 nm), and $Ni_{45}Fe_{55}$ (16 nm) at zero field. Notice that the $B_2^X$ values in the upper layers change from +7 MPa[25] to 0 MPa and -5.7 MPa[26]. For all three bilayers, two power absorption peaks can be seen in Fig. 3(a) with the low-frequency and high-frequency peaks corresponding to the OM and AM, respectively. In the negative-positive configuration of Ni/Ti/FeCoSiB, the OM resonance peak is much stronger than the AM resonance peak, while $Ni_{81}Fe_{19}$/Ti/FeCoSiB and $Ni_{45}Fe_{55}$/Ti/FeCoSiB exhibit an opposite behavior. As illustrated in Fig. 3(b), the enhanced OM in the negative-positive magnetostrictive configuration can be attributed to the opposite driving fields in the bottom and top layers excited by SH-SAWs. Although their magnetizations precess out-of-phase, $h^{A*} \cdot \delta m^A$ always keep in-phase with $h^{B*} \cdot \delta m^B$ due to the antiparallel effective driving fields.

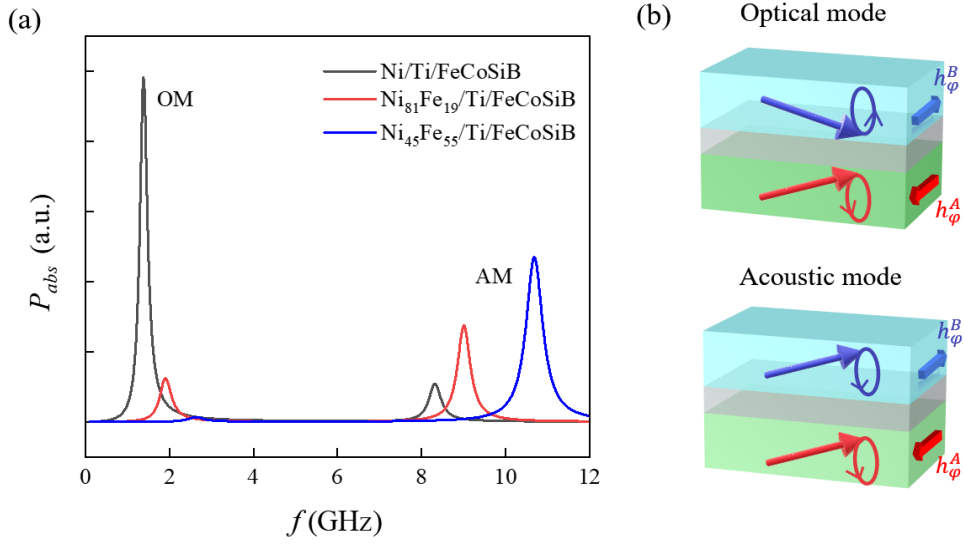

FIG. 3 (a) Calculated normalized SAW power absorption for Ni/Ti/FeCoSiB, $Ni_{81}Fe_{19}$/Ti/FeCoSiB and $Ni_{45}Fe_{55}$/Ti/FeCoSiB configurations. (b) Illustration of optical and acoustic resonance modes for the negative-positive magnetostrictive Ni/Ti/FeCoSiB configuration, where the effective driving fields in the top and bottom layers are always antiparallel.

Since SAWs can only be excited at specific harmonic frequencies for a given wavelength, the resonance frequency of SWs has to be tuned to match that of SAWs by varying the external magnetic fields. To verify the calculation above, we selected the $SH_9$ mode ($k_{SAW} = 3.56 \ \mu m^{-1}$) and measured the field dependent transmission parameters. Although the acoustic insertion loss is large at this frequency, it is mainly caused by the low efficiency of higher-order harmonic modes, and can be improved by narrowing the pitch width of IDTs to excite the fundamental SH mode. The relative change of background-corrected SAW transmission magnitude is defined as $\Delta S_{ij}(H) = S_{ij}(H) - S_{ij}(300\ \text{Oe})$   $(ij = 12\ \text{or}\ 21)$. $S_{ij}(H)$ represents the transmission parameters at SH$_9$ under different external magnetic fields, and $S_{ij}(300\ \text{Oe})$ is the transmission parameters at a fixed field of 300 Oe, which is sufficient to saturate the bilayer.

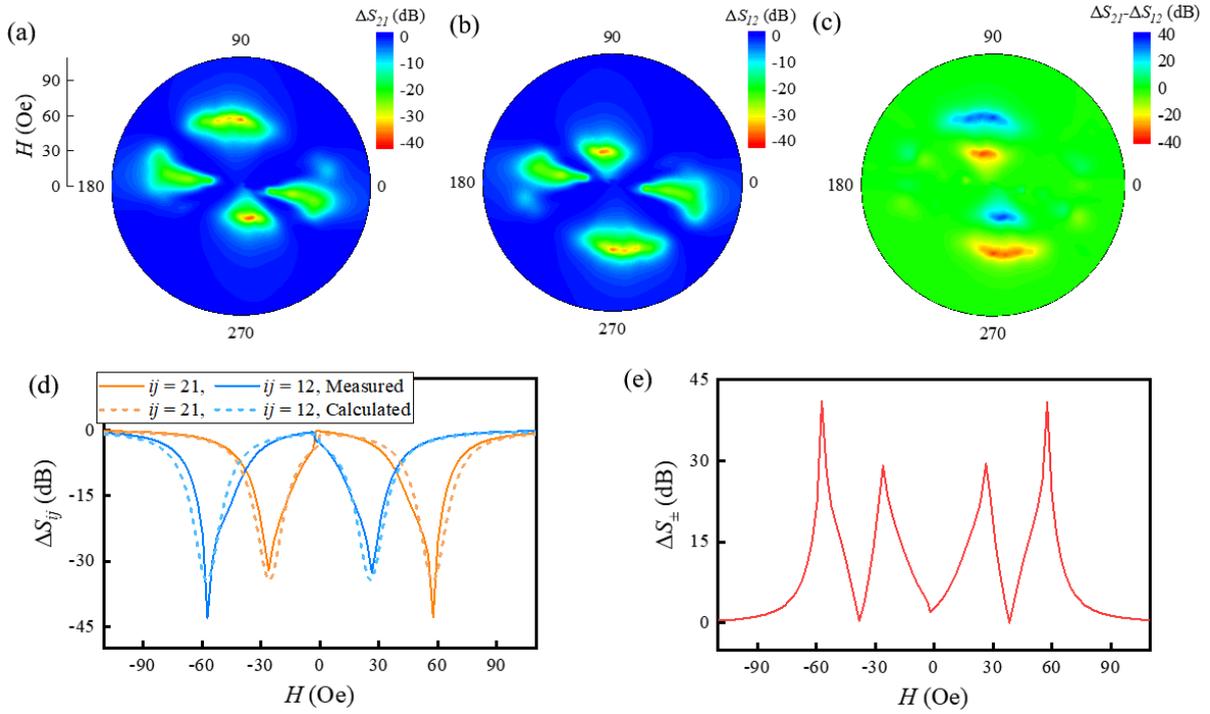

FIG. 4 (a-c) Polar plots of the measured forward ($\Delta S_{21}$), backward ($\Delta S_{12}$) and nonreciprocal transmission ($\Delta S_{21} - \Delta S_{12}$) as a function of applied field $H$ and field angle $\varphi_H$. (d) Measured (solid lines) and calculated (dashed lines) field-dependent $\Delta S_{ij}$ and (e) magnitude nonreciprocity $\Delta S_\pm$ at a fixed $\varphi_H$ of 90°.

Fig. 4(a) and 4(b) show the polar plots of measured $\Delta S_{ij}$ as a function of applied magnetic field for SH$_9$ mode. Strong power absorption is observed for $\varphi_H$ around 0°, 90°, 180°, and 270°, confirming the expected $cos(2\varphi_M)$ angle dependence. Additionally, comparing Fig. 4(a) and 4(b), one can find large variations in the corresponding SWR fields near 90° and 270°. Generally, the nonreciprocity of SAW transmission is defined as $\Delta S_{21} - \Delta$ $S_{12}$. As shown in Fig.4(c), the strongest nonreciprocity (in other words, isolation coefficient) exceeds 40 dB near $\varphi_H = 90°$.

Based on our recently developed dynamic magnetoelastic coupling model[27,28], $\Delta S_{ij}$ can be calculated using the formula below

$$\Delta S_{ij} = \exp(k_I(H)L)\exp(-jk_R(H)L), \ ij = 12, 21. \tag{5}$$

where $k(H) = 2\pi f_{SAW}/v(H)$ is the magnetoacoustic wave vector, and $v(H)$ is the field-dependent magnetoacoustic wave velocity. $k_R(H)$ and $k_I(H)$ are the real and imaginary part of $k(H)$, respectively (see Section B of supplementary material). Thus, the magnitude of $\Delta S_{ij}$ can be written as

$$|\Delta S_{ij}| = 20\log_{10}(\exp(k_I(H)L)). \tag{6}$$

Both measured and calculated field-dependent $\Delta S_{ij}$ results for a fixed field orientation of $\varphi_H = 90°$ are further shown in Fig.4(d), demonstrating good consistency.

Under zero field, the OM frequency is lower than that of SH$_9$ mode, so there is no coupling between SWs and SAWs. At $H_{ref1} = 26$ Oe, $\Delta S_{12}$ reaches its minimum value of -33 dB. This indicates that $f_{OM}(-k)$ can intersect with $f_{SAW}$ along the backward propagation direction, resulting in significant power absorption. Since there is a large $\Delta f_{OM}$ according to Fig. 2(b), $f_{OM}(+k)$ is still lower than $f_{SAW}$ along the forward direction, therefore, causing a large magnitude nonreciprocity ($\Delta S_\pm = |\Delta S_{21} - \Delta S_{12}|$) of 30 dB (or 60 dB/mm), as shown in Fig. 4(e). A second SWR appears at $H_{ref2} = 57$ Oe, and is accompanied by an even more pronounced SAW power absorption. Compared to the previous resonance, a larger applied field causes that $f_{OM}(+k)$ and $f_{SAW}$ cross with each other along the forward propagation direction ($\Delta S_{21} = -43$ dB), while $f_{OM}(-k)$ is now higher than $f_{SAW}$, thus exhibiting weak power absorption along the backward direction ($\Delta S_{12} = -2$ dB). As shown in Fig. 4(e), the magnitude nonreciprocity of the second SWR reaches 41 dB (or 82 dB/mm). These values are comparable with those reported in synthetic antiferromagnetic systems[15,16,20] at such a low frequency, although the magnetization in the neighboring layers are parallel with each other.

Our results open a way to develop efficient, very stable microwave SAW isolators. Isolators, however, are not the only microwave frequency nonreciprocal devices, that are in demand for SAW-based signal processing technology. Circulators and nonreciprocal phase shifters are also needed. A standard circulator scheme relies on the phase nonreciprocity, i.e. the phase accumulation for waves propagating in opposite direction is not the same. Previously, Verba et al.[29] proposed a SAW circulator, in which three nonreciprocal phase shifters based on SAFM heterostructures are combined in a SAW ring resonator. Next, we will show that the negative-positive magnetostrictive bilayer can also support nonreciprocal π-phase accumulation.

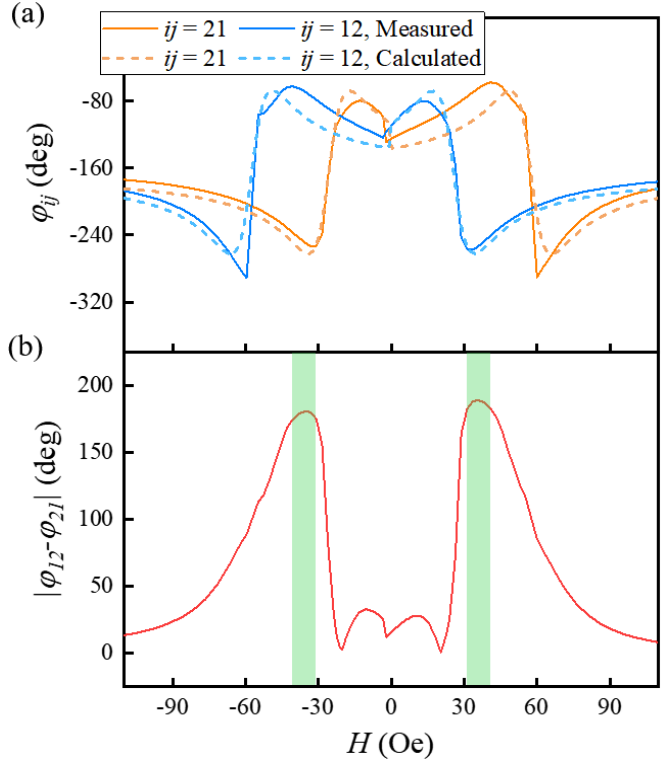

FIG. 5 (a) SAW transmission phase ($\varphi_{ij}$) and (b) phase nonreciprocity ($|\varphi_{21} - \varphi_{12}|$) as a function of applied field for a given $\varphi_H$ of 90°. The shaded area in (b) corresponds to the field range of $|\varphi_{21} - \varphi_{12}| \geq 180°$.

For the Ni/Ti/FeCoSiB bilayer structure, we can extract the SAW transmission phase of the delay line for a fixed field angle of $\varphi_H = 90°$ using the measured $S_{ij}(H)$ and the formula below:

$$\varphi_{ij} = \arctan\left(\frac{\text{Im}(S_{ij})}{\text{Re}(S_{ij})}\right), \qquad (7)$$

as shown in Fig. 5(a). Clearly, $\varphi_{ij}$ exhibits sharp up and down at the SWR fields corresponding to the peaks of $\Delta S_{ij}$ (Fig. 4d). We have also calculated the phase of the delay line using

$$\varphi_{ij} = -k_R(H)L \qquad (8)$$

Again, the calculated results are in good accordance with the measured ones. Strong phase nonreciprocity exceeding 180° is observed in Fig. 5(b) in the field range of ±(31-42) Oe, as highlighted by the shaded area. In particular, $|\varphi_{21} - \varphi_{12}|$ reaches 188° (376°/mm), which is desired for SAW circulators. Moreover, the field range of the π-phase nonreciprocity does not overlaps with those SWR fields, but is lower than $H_{ref2}$ and higher than $H_{ref1}$. Therefore, a relatively low propagation loss $\Delta S_{ij} = 9.1$ dB is obtained near 36 Oe.

However, a practical SAW circulator requires both π-phase nonreciprocity and low propagation loss below 3 dB. According to Ref. [15,29], the necessary conditions to fulfill these criteria are a sufficiently large nonreciprocal splitting of the SW dispersion, $\Delta f_{OM} \gg \Delta f_{me} \gg \sqrt{\Gamma_{SAW}\Gamma_{SW}}/2\pi$, and strong MEC larger than the damping rate of the SWs, $\Delta f_{me} > \Gamma_{SW}/2\pi$. Here, $\Delta f_{me}$ represents the magnetoelastic coupling strength, $\Gamma_{SAW}$ and $\Gamma_{SW}$ are the damping rate of SAWs and SWs, respectively. In our device, for the 57 Oe applied field, we obtain $\Delta f_{OM} = 482$ MHz, $\Delta f_{me} = 52$ MHz, $\Gamma_{SAW} = 2\pi \times 0.56$ MHz and $\Gamma_{SW} = 2\pi \times 182$ MHz (see Section C of supplementary material). These results are comparable to those determined in FeGaB(20 nm)/Al$_2$O$_3$(5 nm)/FeGaB(20 nm) at 1.435 GHz[15], where $\Delta f_{OM} = 218$ MHz, $\Delta f_{me} = 12$ MHz, $\Gamma_{SAW} = 2\pi \times 30$ kHz. But the $\Gamma_{SW}$ of Ni/Ti/FeCoSiB is higher than that of FeGaB/Al$_2$O$_3$/FeGaB, $\Gamma_{SW} = 2\pi \times 90$ MHz. Although the first condition of $\Delta f_{OM} \gg \Delta f_{me} \gg \sqrt{\Gamma_{SAW}\Gamma_{SW}}/2\pi$ is satisfied, the second condition is still not fulfilled due to the relatively large $\Gamma_{SW}$.

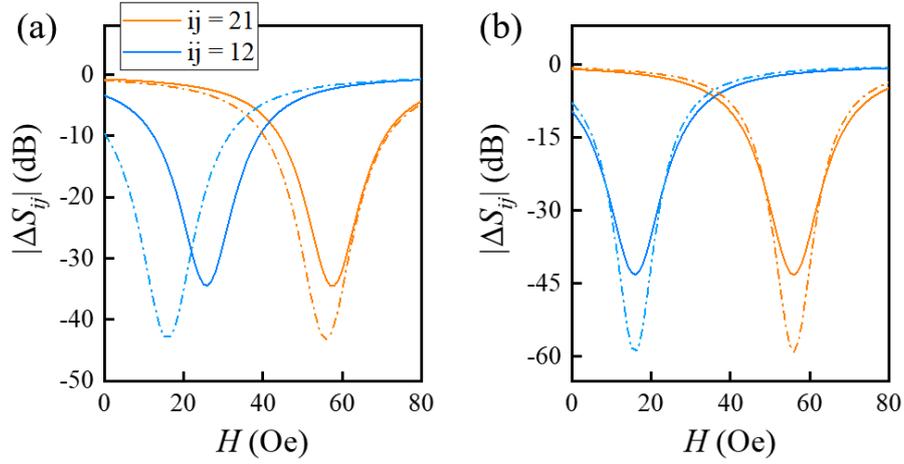

FIG. 6 (a) Calculated $|\Delta S_{ij}|$ of Ni/Ti/FeCoSiB with different layer thicknesses of 16 nm (solid line) and 20 nm (dot-dashed line), and (b) calculated $|\Delta S_{ij}|$ of Ni(20 nm)/Ti/FeCoSiB(20 nm) with different damping factors of Ni 0.012 (solid line) to 0.008 (dot-dashed line). The damping factor of FeCoSiB is 0.008.

There are some useful measures that could be carried out to further optimize the negative-positive magnetostrictive bilayer structure. Firstly, according our phenomenological model in Sec. A of supplementary material, one can increase $\Delta f_{OM}$ to enlarge $\Delta H_{ref}$. Fig. 6(a) shows the calculated $|\Delta S_{ij}|$ upon changing the thickness of the Ni and FeCoSiB layers from 16 to 20 nm. As can be seen, $\Delta H_{ref}(= H_{ref2} - H_{ref1})$ increases from 31 to 40 Oe, which helps that the magnitude nonreciprocity increases 8 dB and the insertion loss decreases 0.6 dB. Secondly, reducing the Gilbert damping in the magnetic bilayer can also lower insertion loss. When we change the damping factor of Ni layer from 0.012 to 0.008, the magnitude nonreciprocity is further improved to 57 dB, meanwhile the insertion loss is reduced down to -1.3 dB, as shown in Fig. 6(b). Notice that FeCoSiB has a measured low damping factor of 0.008[30]. So, the key is to develop negatively magnetostrictive film with a low Gilbert damping factor. For example, doping non-magnetic elements in nickel to form an amorphous phase may reduce the Gilbert damping, although there is currently limited research on this. Finally, utilizing TF-SAW technology, such as bonding LiTaO$_3$ on a high-velocity substrate[31], is beneficial to better confine the acoustic wave to the substrate surface, therefore increasing $\Delta f_{me}$.

In summary, we demonstrated simultaneous large magnitude and phase nonreciprocity of SAWs up to 82 dB/mm and 376°/mm using a magnetoacoustic hybrid device integrated with a negative-positive magnetostrictive Ni/Ti/FeCoSiB bilayer. Our theoretical calculations based on the dynamic magnetoelastic coupling model are in good agreement with experimental results. In spite of the parallel magnetization of the two layers, SH-SAW excitation can greatly enhance optical-mode spin wave resonance but suppress the acoustic-mode ones owing to the opposite effective driving fields. The relatively lower frequencies of optical mode migrate the requirement on nanolithography, while the negative-positive magnetostrictive bilayer structure provides more freedom on structural design than synthetic antiferromagnets. In our opinion, the propagation loss can be improved by optimizing the thickness of the negative-positive magnetostrictive bilayer structure and employing negative magnetostrictive film with a low Gilbert damping factor.

**Supplementary Material**

See the supplementary material for extensive discussion of the theoretical computation method of SW dispersion and SAW transmission characteristics mentioned in the paper.

**Acknowledgement**

This work is supported by the National Natural Science Foundation of China (Grant No. 61871081 and 61271031) and the Natural Science Foundation of Sichuan Province under Grant No. 2022NSFSC0040.

**Data availability**

The data that support the findings of this study are available within the article and its supplementary material.